
\magnification=\magstep1
\baselineskip=20pt
\centerline{The Ptolemaic Gamma-Ray Burst Universe}
\bigskip
\centerline{J. I. Katz}
\medskip
\centerline{Department of Physics and McDonnell Center for the Space
Sciences}
\centerline{Washington University, St. Louis}
\vfil
\noindent
2 Copies, 10 pp., 0 figures, 0 tables \par
\noindent
Mailing address: Department of Physics, Washington University, St. Louis,
Mo. 63130 \par
\noindent
Telephone: 314-935-6202 \par
\noindent
e-mail: katz@wuphys.wustl.edu \par
\eject
\centerline{Abstract}
\bigskip
The BATSE experiment on GRO has demonstrated the isotropic arrival
directions and flat $\log N$ {\it vs.} $\log S$ of cosmic gamma-ray bursts.
These data are best explained if the burst sources are distributed throughout
an extended spherical Galactic halo, as previously suggested by Jennings.
The halo's radius is at least 40 Kpc, and probably is more than 100 Kpc.  I
consider possible origins of this halo, including primordial formation and
neutron stars recoiling
from their birthplaces in the Galactic disc.  A simple geometrical model
leads to a predicted relation between the dipole and quadrupole anisotropy.
I suggest that neutron stars born with low recoil become millisecond
pulsars, while those born with high recoil become the sources of gamma-ray
bursts; these populations are nearly disjoint.  Quiescent counterparts of
gamma-ray bursts are predicted to be undetectably faint.
\par
\vfil
\eject
The first results from the BATSE on GRO (BATSE Science Team 1991)
have revived the question of the distribution of gamma-ray burst sources in
space.  Their chief results, isotropy of gamma-ray burst directions and a
$\log N$ {\it vs.} $\log S$ slope significantly flatter than -1.5, confirm
earlier reports (see, for example, Meegan, Fishman and Wilson 1985 and the
review by Cline 1984).  Questions of relative calibration of different
instruments and the paucity of good directional data permitted skepticism in
the past.  Such skepticism is no longer tenable, and the
theoretical questions raised earlier must be faced.

An isotropic distribution of sources implies that, out to the maximum
distance of observation permitted by instrumental sensitivity, all
directions contain equivalent source populations.  The source population
for an observed flux or fluence $S$ is expressed as the integral
$$N({\hat \Omega}, S) = \int n({\hat \Omega}, r, 4 \pi r^2 S) 4 \pi r^2\,
dr, \eqno(1)$$
where $n({\hat \Omega}, r, 4 \pi r^2 S)$ is the density of sources at a
distance $r$ from the observer in the direction $\hat \Omega$ radiating a
total power or energy $4 \pi r^2 S$; a temporal average is implicit, and it
is assumed that the sources radiate isotropically.  While it is possible
to obtain $N$ independent of $\hat \Omega$ for a variety of artificial
spatial distributions $n$, the only plausible $n$ which does not require the
observer to be in a preferred position (for example, at the center of
spherical shells) is a homogeneous distribution out to the greatest
distances at which the most luminous burst may be observed.  Such a
distribution implies $N \propto S^{-3/2}$, in contradiction to observation.
This dilemma was fully appreciated and discussed by several of the
contributors to the 1983 Symposium on High Energy Transients in Astrophysics
(Woosley 1984).

The only geometric resolution of this problem is the Ptolemaic gamma-ray
burst universe, as discussed by Cline (1984).  The flat dependence of $N$ on
$S$ (equivalent to a peak in a $V/V_{max}$ distribution at $V \ll V_{max}$,
and to a deficiency of faint bursts) is explained if the sources are
distributed throughout a volume of
finite extent, with most bursts within the volume luminous enough to
be observed.  This volume is spherical, with the Earth at its center.  The
sources must be at cosmological distances, or be distributed in an extended
halo around our Galaxy (such as the Massive Extended Halo suggested by
Jennings 1984), or be evidence of a previously unsuspected Earth-centered
spherical distribution.

At cosmological distances a typical burst of $10^{-7}$ erg/cm$^2$ fluence
must have radiated $\sim 10^{50}$ erg in gamma-rays.  This is
energetically consistent with the spiraling-in or coalescence of binary
neutron stars, if they convert gravitational
energy to gamma-rays, perhaps through a magnetohydrodynamic dynamo, or with
a relativistic supernova shock breakout, but the mechanisms
are obscure.  The breakout of a relativistic shock from a
spherical shell of radius $R_s$ would produce a pulse of radiation whose
envelope would be proportional to the local (relativistically collimated)
emission pattern $f(\theta(t))$, where $\theta(t) = (2ct/R_s)^{1/2}$ and
$\theta \sim \gamma^{-1} \ll 1$ is assumed.  The rapid fluctuations of
intensity under a flat envelope observed in many bursts might then be explained
by shock refraction in a turbulent stellar envelope.  Unfortunately, the
implied values of $R_s \sim 2ct_b \gamma^2$, where $t_b$ is the burst
duration, are unreasonably large, particularly because the $\gamma$-$\gamma$
pair-production constraint on 100 MeV gamma-rays implies $\theta < 0.01$.

Such catastrophic events might be expected all to have similar
time histories, or to fall into a small number of homogeneous classes,
inconsistent with observation.  The observed soft gamma repeaters, the 8
second periodicity of the March 5, 1979 burst, the absence of bursts from
local supernovae, the identification of 400--500 KeV
spectral features with positron annihilation lines of modest redshift, and
the complex and varied time histories of intensity within bursts
all disfavor these cosmological hypotheses.

If an observer is displaced by a distance $a$ from the center of a sphere of
radius $R \gg a$, the mean cosine of the angle $\theta$ between the
points in the sphere and the direction of displacement is found by
integration, keeping only the lowest order terms in $a/R$:
$$\langle \cos \theta \rangle = {3 \over 4 \pi R^3} \int\int\int
d\phi \sin\theta_0 d\theta_0 r^2 dr \cos\theta \approx {a \over R},
\eqno(2)$$
where $\theta = \theta_0 + \Delta \theta \approx \theta_0 - {a \over r}
\sin\theta_0$ and the integration runs over coordinates $(r, \theta_0,
\phi)$ measured from the center of the sphere.  Adopting $a = 8.5$ Kpc for our
displacement from the center of the Galaxy and $\langle \cos\theta \rangle <
0.08$ (BATSE 1991; $1 \sigma$ bound) implies $R > 106$ Kpc; the $3 \sigma$
bound of $\langle \cos\theta \rangle < 0.22$ relaxes this to $R > 39$ Kpc.
There is little other direct evidence for a spherical component of our Galaxy
extending to these radii, but it is believed (based on dynamical
measurements) that the mass distributions of many galaxies similarly extend
far beyond their visible radii.  The requirement that the halo be spherical
constrains models of its formation and gravitational potential.

It is possible that the objects (by consensus, probably neutron stars)
producing gamma-ray bursts were formed in this extended halo early in the
history of the Galaxy, and that they remain active for $\ge 10^{10}$ years.
The total mean gamma-ray burst flux crossing the Galactic plane is $\sim
10^{-11}$ erg/cm$^2$sec, corresponding to a Galactic luminosity of $\sim
10^{37}$ erg/sec even for a very extended halo ($R \sim 100$ Kpc), or to
$\sim 10^{26}$ erg/sec $M_\odot$ for a halo mass of $\sim 10^{11} M_\odot$,
comparable to the mass of the visible Galaxy.  The cumulative power radiated
is $\sim 3 \times 10^{43}$ erg/$M_\odot$.  If all the mass were in neutron
stars, this energy would be comparable to their magnetospheric energy for
$B \sim 10^{13}$ gauss, or to their rotational energy with moments of
inertia of $10^{45}$ gm cm$^2$ and spin periods $\sim 25$ seconds.
Presumably only a fraction of the mass is in neutron stars; their expected
magnetospheric energy is inadequate, but their rotational energy may be
sufficient.

Alternatively, the sources of gamma-ray bursts may be born within the
visible Galaxy but escape to an extended halo.  It is well known (Lyne,
Anderson and Salter 1982) that many (but probably not all) pulsars are
produced with recoil velocities $\sim 200$ km/sec, sufficient to escape the
Galactic disc.  When combined with their initial orbital kinetic energy,
they may be sufficiently energetic to escape into an extended halo.
The time required to fill a halo of radius $R \sim$ 40--100 Kpc is 1--3
$\times 10^8$ years.  This hypothesis is complicated by the fact that their
initial orbital velocities about the Galactic center will tend to concentrate
the neutron stars to low Galactic latitudes, and any low-recoil fraction
will remain in the Galactic plane, concentrated toward the Galactic center.

The lifetimes of these neutron stars as gamma-ray burst
sources must not be less than this filling time (or the bursts would be
concentrated within the visible Galaxy, and peak towards its center), but
could be arbitrarily greater.  It is not possible to estimate the number of
neutron stars required because their burst frequency is unknown.
If the neutron stars freely escape to
infinity and radiate indefinitely their $n \propto r^{-2}$ implies an
asymptotic $d\log N / d\log S = -0.5$, not as flat as the  $d\log N / d\log S
= 0$ found for a confined space distribution, but flat enough to explain the
data.  The energy required is the same as in the case of gamma-ray
burst sources born in the halo.

For a simple model of gamma-ray burst sources recoiling from their
birthplaces in the Galactic disc, it is possible to calculate the expected
dipole moment of their distribution of arrival directions with respect to
the Galactic center and the quadrupole moment with respect to the Galactic
plane.  Assume that we reside a distance $r_0$ from the center of a Galactic
disc of radius $L$, uniformly populated with source birthplaces, and that
each point in the disc uniformly fills with observable gamma-ray bursts a
sphere of radius $R \gg L,\, r_0$ with itself at its center.
The uniformly filled sphere is a rough
approximation to the motion of neutron stars in the unknown potential of the
Galactic halo.  Using the result (2), which now describes the observed
asymmetry of each sphere, with $a^2 = r^2 + r_0^2 -2rr_0 \cos\phi$ and
$(r,\phi)$ denoting the coordinates of the disc point, projecting this
asymmetry onto the direction to the Galactic center, and integrating over
the disc yields
$$\langle \cos \theta \rangle \approx {r_0 \over R}, \eqno(3)$$
where $\theta$ is the angle between a burst's arrival direction and that to
the Galactic center.  Application to the BATSE data yields the same bounds
on $R$ as discussed above, although now $R$ refers to the radius of a sphere
populated by gamma-ray bursts from a given neutron star birthplace, rather
than the radius of a sphere of source positions centered at the Galactic
center; the distinction is slight for $R \gg L,\, r_0$, as must be the case.
If the assumption of uniform filling is replaced by $d^{-2}$ density
distribution, where $d$ is the distance from the birthplace (reflecting an
outward streaming distribution, rather than a trapped one), then the result
(3) is multiplied by a factor $O(\ln(R/r_0)) \sim 2$.

It is similarly possible to calculate an average $\langle \sin^2\beta
\rangle$, where $\beta$ is a burst's Galactic latitude.  For an observer in
the Galactic equatorial plane but at a distance $a$ from the center of a
uniformly populated sphere of radius $R$,
$$\langle \sin^2\beta \rangle \approx {1 \over 3} - {4 a^2 \over 5 R^2},
\eqno(4)$$
to lowest non-vanishing order in $a/R$.  Averaging over spheres centered in
the Galactic disc yields
$$\langle \sin^2\beta \rangle \approx {1 \over 3} - {4 r_0^2 + 2 L^2 \over 5
R^2}. \eqno(5)$$
Because of the quadratic dependence on $R$ the resulting bounds are less
strict than those obtained from (2) or (3).

The results (3) and (5) may be combined to predict a relation between these
two angular averages, if the relation between $L$ and $r_0$ is known.
Take, for simplicity, the reasonable estimate $L = \sqrt{2} r_0$.  Then
$$\langle \sin^2\beta \rangle \approx {1 \over 3} - {8 \over 5} \langle
\cos\theta \rangle^2. \eqno(6)$$
This relation offers a test of the simple geometrical model used here.  This
test is readily satisfied by the extant BATSE data, but may become more
discriminating as more data accumulate.

The gamma-ray burst of March 5, 1979 may, in part, be explained by the
extended halo hypothesis.  If most other bursts are at distances less than
the thickness of the Galactic disc, as was often supposed before the BATSE
data became available, then the March 5, 1979 event was extraordinary,
because its luminosity must have been $\sim 10^5 - 10^6$ times greater than
that of a typical burst.  It is hard to reconcile this interpretation with
the absence of bursts similar to that of March 5, 1979, but occurring in
our own Galaxy, which contains $\sim 10$ times the mass of the LMC, and
might therefore be expected to have $\sim 10$ times as many such events
(but closer and even brighter!).  If, instead, bursts in general are
distributed throughout a very extended halo of our Galaxy, extending to
distances at least as great as that of the LMC, then the March 5, 1979 burst
would be explicable as an unusually (but not extraordinarily) luminous
member of a class of events with a broad distribution of luminosities.

Another consequence of the hypothesis of the birth of gamma-ray burst
sources in the Galactic disc is a further development of the empirical
division of neutron star births into high and low recoil classes (Katz
1975).  Millisecond pulsars, found in superabundance in dynamically fragile
globular clusters (like their probable ancestors, the X-ray binaries) must
have been formed recoillessly.  Slowly spinning pulsars, most of which are
single and are observed (Lyne, Anderson and Salter 1982) to have high space
velocities (and hence high recoil) are the ancestors of gamma-ray burst
sources.

These categories are nearly disjoint because low-recoil millisecond pulsars
retain their companions, initially suppressing their pulsar emission
(Shvartsman 1971), but later spinning them up.  The observed small magnetic
dipole moment $\mu$ of millisecond pulsars reduces the power $<\mu^2 c/
(3 R_{ns}^2)$ potentially released by magnetospheric dissipation in a
gamma-ray burst, explaining the absence of this Galactocentrically-concentrated
population from the spatial distribution of bursts, although their large
rotational energy suggests that a significant luminosity in microbursts
may be possible.

High-recoil neutron stars, born with slow spin but which lose any binary
companion, must retain their magnetic fields for a very long time in order
to produce gamma-ray bursts (empirically, Her X-1 has retained its
large field for at least $3 \times 10^7$ years, the time required for it to
reach its height above the Galactic plane, and perhaps much longer).
Low-recoil single neutron stars and high-recoil neutron stars which retain
their companions, although individually interesting (Her X-1, PSR1913+16),
are uncommon.

If gamma-ray bursts are distant then they must be very powerful.
At 100 Kpc a strong $10^{-5}$
erg/cm$^2$ burst, radiating isotropically, emits $10^{43}$ erg, posing
energetic problems similar to those posed by the March 5, 1979 event (Katz
1982).  Nonthermal mechanisms involving the flare-like dissipation of much
of a magnetosphere's energy are an attractive explanation of much gamma-ray
burst phenomenology, but they are barely energetically adequate for a single
burst.  The possibility of regeneration of magnetospheric energy even during
a burst (perhaps by twisting up the magnetosphere by differential rotation)
must be considered.  Note, though, that even if
the burst population fills a sphere of radius $R$ the brighter bursts will,
on average, be significantly closer than $R$, reducing their energy
requirements but increasing the spatial anisotropy of their distribution.

A final consequence of the distances suggested here is that quiescent
burst counterparts will be too faint to observe.  Stellar companions
(unless close and compact) will have been lost when recoil was imparted.
Accurate burst coordinates and deep searches will show only blank fields, as
has been the case for the few accurate positions presently available
(Schaefer 1990).  An essentially distance-independent argument (Katz 1985)
implies that Schaefer's (1990; but see \.Zytkow 1990) reported optical
flashes can only be explained as nonthermal emission.  At a distance of 100
Kpc thermal reprocessing would be inadequate to explain the reported
intensities (the inferred absolute magnitudes would be brighter than -15!).
Similarly, at these distances self-absorption suppresses the reprocessed
X-ray flux, explaining the low X-ray luminosity of gamma-ray bursts (Imamura
and Epstein 1987).

I thank G. J. Fishman for supplying data in advance of publication and F. J.
Dyson for discussions.
\par
\vfil
\eject
\centerline{References}
\bigskip
\parindent=0pt
BATSE Science Team 1991, I. A. U. Circ. 5358. \par
\medskip
\hangindent=20pt
\hangafter=1
Cline, T. L. 1984 in {\it High Energy Transients in Astrophysics}, ed. S. E.
Woosley (AIP: New York) p. 333.
\medskip
Imamura, J. N., and Epstein, R. I. 1987, {\it Ap. J.} {\bf 313}, 711.
\medskip
\hangindent=20pt
\hangafter=1
Jennings, M. C. 1984, in {\it High Energy Transients in Astrophysics}, ed.
S. E. Woosley (AIP: New York) p. 412.
\medskip
Katz, J. I. 1975, {\it Nature} {\bf 253}, 698.
\medskip
Katz, J. I. 1982, {\it Ap. J.} {\bf 260}, 371.
\medskip
Katz, J. I. 1985, {\it Ap. Lett.} {\bf 24}, 183.
\medskip
Katz, J. I. 1987, {\it High Energy Astrophysics} (Addison-Wesley: Menlo
Park) \S{\bf 4.5.1}.
\medskip
Lyne, A. G., Anderson, B., and Salter, M. J. 1982, {\it M. N. R. A. S.} {\bf
201}, 503.
\medskip
Meegan, C. A., Fishman, G. J., and Wilson, R. B. 1985, {\it Ap. J.} {\bf
291}, 479.
\medskip
Schaefer, B. E. 1990, {\it Ap. J.} {\bf 364}, 590.
\medskip
Shvartsman, V. F. 1971, {\it Sov. Astron.--AJ} {\bf 15}, 342.
\medskip
Woosley, S. E. 1984, ed. {\it High Energy Transients in Astrophysics} (AIP:
New York).
\medskip
\.Zytkow, A. 1990, {\it Ap. J.} {\bf 359}, 138.
\vfil
\eject
\bye
\end